\documentclass[12pt,preprint]{aastex}

\newcommand{\be}{\begin{equation}}
\newcommand{\ee}{\end{equation}}
\begin{document}

\title{GRAVITATIONAL COLLAPSE OF MAGNETIZED CLOUDS \\ II. THE ROLE OF OHMIC DISSIPATION}

\author{Frank H. Shu}
\affil{Physics Department, National Tsing Hua University \\
              Hsinchu 30013, Taiwan, Republic of China \\
              shu@mx.nthu.edu.tw}

\author{Daniele Galli}
\affil{INAF-Osservatorio Astrofisico di Arcetri \\
              Largo Enrico Fermi 5, I-50125 Firenze, Italy \\
              galli@arcetri.astro.it}

\author{Susana Lizano}
\affil{Centro de Radioastronom\'{\i}a y Astrof\'{\i}sica, UNAM \\
              Apdo. Postal 3-72, Morelia, Michoac\'an 58089, Mexico\\
              s.lizano@astrosmo.unam.mx}

\author{Mike Cai}
\affil{Physics Department, National Tsing Hua University \\
              Hsinchu 30013, Taiwan, Republic of China \\
              mike@phys.nthu.edu.tw}

\begin{abstract}
We formulate the problem of magnetic field dissipation during the
accretion phase of low-mass star formation, and we carry out the first
step of an iterative solution procedure by assuming that the gas is in
free-fall along radial field lines.  This so-called ``kinematic
approximation'' ignores the back reaction of the Lorentz force on the
accretion flow. In quasi steady-state, and assuming the resistivity
coefficient to be spatially uniform, the problem is analytically
soluble in terms of Legendre's polynomials and confluent hypergeometric 
functions.  The dissipation of the magnetic field occurs inside a
region of radius inversely proportional to the mass of the central star
(the ``Ohm radius''), where the magnetic field becomes asymptotically
straight and uniform.  In our solution, the magnetic flux problem of
star formation is avoided because the magnetic flux dragged in the
accreting protostar is always zero. Our results imply that the
effective resistivity of the infalling gas must be higher by
several orders of magnitude than the microscopic electric resistivity, to
avoid conflict with measurements of paleomagnetism in meteorites and
with the observed luminosity of regions of low-mass star formation.
\end{abstract}

\keywords{ISM:clouds --- ISM: magnetic fields --- magnetohydrodynamics ---
planetary systems: protoplanetary disks --- stars: formation}

\section{Introduction}

{If the magnetic field of a collapsing interstellar gas cloud
remained frozen in the gas, the resulting surface magnetic field of the
newborn protostar would exceed observed stellar fields by almost four
orders of magnitude (the so-called ``magnetic flux problem'', see e.g.,
Mestel \& Spitzer~1956). It follows that the excess magnetic flux of a
cloud must be dissipated at some stage during the process of star
formation.  Assuming ideal magnetohydrodynamic (MHD) conditions,} Galli
et al.~(2006, hereafter Paper I) obtained an analytical solution for
the inner regions of an isothermal, magnetized, rotating cloud
undergoing gravitational collapse and showed that the long-lever arms
of the strong field trapped in the central protostar in a
split-monopole configuration would cause so much magnetic braking as to
make impossible the formation of a centrifugally supported disk around
the central object (see also Allen et al. 2003a,b). {Thus, the
dissipation of magnetic field must occur prior or simultaneosly to the
formation of a circumstellar disk.} In the current paper, we remove the
assumption of field freezing by allowing the non-ideal effect of finite
electric resistivity to operate, {and we determine the value of
the resistivity coefficient required to solve the magnetic flux problem
during the accretion phase of low-mass star formation.

The paper is organized as follows: in Section~2 we formulate the basic
equations of the problem; in Section~3 we estimate the value of the
resistivity coefficient in the central region of collapse from the
available observational constraints; in Section~4 we check the validity
of our approximations; in Section~5 we compare our results with those
of previous works; finally, in Section~6, we summarize our
conclusions.  The reader not interested to the mathematical derivation
of our results, can go directly to eq.~(\ref{sol}).}

\section{Formulation of the problem}

{Neglecting the effects of ambipolar diffusion (but see Section~6),
the evolution of the magnetic field in a collapsing cloud is governed
by} the induction equation with ohmic dissipation,
\be
{\partial {\bf B}\over\partial t}+\nabla\times({\bf
B}\times {\bf u})= -\nabla\times (\eta\nabla\times {\bf B}),
\label{ind1}
\ee
where $\eta$ is the coefficient of ohmic
resistivity, related to the electric conductivity $\sigma$ by
\be
\eta={c^2\over 4\pi\sigma}.
\label{cond}
\ee
Assuming axial
symmetry, and a purely poloidal magnetic field, we can ``uncurl''
eq.~(\ref{ind1}) and express it in scalar form as
\be
{\partial \Phi\over\partial t}+{\bf u}\cdot\nabla\Phi=\eta {\cal S}(\Phi),
\ee
where $\Phi$ is the flux function, defined in spherical
coordinates by
\be
{\bf B}=\nabla\times \left[{\Phi(r,\theta)
\over 2\pi r\sin\theta} \hat{\bf e}_\varphi \right],
\label{defB}
\ee
and ${\cal S}$ is the Stokes operator
\be
{\cal S}\equiv
{\partial^2\over\partial r^2}+{1\over
r^2}{\partial^2\over\partial\theta^2} -{\cot\theta\over
r^2}{\partial\over\partial\theta}.
\ee
Since $B_r$ is antisymmetric with respect to the midplane,
eq.~(\ref{defB}) implies that the flux function $\Phi$ is even with
respect to $\theta=\pi/2$.  In addition, because the field is finite on
the polar axis (except at the origin), the flux function $\Phi$ must
vanish for $\theta=0$ and $\theta=\pi$ \footnote{The physical flux
threading a circle of radius $r$ on the midplane is given by the value
$\Phi(r, \theta=\pi/2)$.}

Following Paper I, we assume that the velocity is given by free-fall on
a point mass $M_\star$ at the origin,
\be 
{\bf u}(r)=-\left({2GM_\star\over r}\right)^{1/2}\hat{\bf e}_r,
\label{freefall}
\ee
and we ignore the reaction of the magnetic field on the flow (the
validity of this assumption is to be checked a posteriori).
Because the dissipation is likely to occur in a region of space
across which the flow time is short in comparison with the
evolutionary time of the collapse of the molecular cloud core, we
may look for a quasi-steady state in which the advection of
magnetic flux by the flow is balanced by ohmic dissipation,
\be
-\left({2GM_\star\over r}\right)^{1/2}{\partial\Phi\over\partial
r} =\eta {\cal S}(\Phi).
\label{ss}
\ee
At large $r$, the flux function must asymptotically approach that of a
split monopole,
\be
\lim_{r\rightarrow\infty}\Phi(r,\theta)=\Phi_\star(1-|\cos\theta|),
\label{bc}
\ee
where
\be
\Phi_\star=2\pi\lambda_\star^{-1} G^{1/2} M_\star
\label{finf}
\ee
is the flux trapped by the central protostar, and the parameter
$\lambda_\star$ is the mass-to-flux ratio of the central split monopole
in non-dimensional units (see Paper~I). {Note that $\lambda_\star$
is a measure of the magnetic flux trapped in the star under ideal MHD
condition, and does not correspond to the actual mass-to-flux ratio
measured in young stars ($\sim 10^3$--$10^4$ in the same units). In
Paper I, we obtained $1 \la \lambda_{\star} \la 4$ by
connecting the analytic inner collapse solution to the ideal MHD
numerical models of Allen et al.~(2003a,b).  In the following, we will
use $\lambda_\star \approx 2$ as a fiducial value.}

\subsection{Nondimensional variables}

To be specific and for simplicity, we assume that $\eta$ is
spatially constant and we define the nondimensional variables
\be
r=r_{\rm Ohm} x, \qquad \Phi(r,\theta)=\Phi_\star\phi(x,\mu),
\ee
where $\mu=\cos\theta$ and
\be
r_{\rm Ohm}\equiv {\eta^2\over 2GM_\star}
\label{ohmrad}
\ee
is the {\it Ohm radius}.  We may then justify the assumption of a
quasi-steady state if the fractional variations of the parameters of
the problem, $M_\star$ and $\Phi_\star$, are negligible over a diffusion time
\be
t_{\rm Ohm} \equiv {r_{\rm Ohm}^2 \over \eta} =
{\eta^3 \over 4G^2M_\star^2}.
\label{tOhm}
\ee
{ Because advection is balanced against diffusion, the expression
(\ref{tOhm}) is also the time it takes to cross $r_{\rm Ohm}$ at the
local free-fall velocity
\be
u(r_{\rm Ohm})=\left({2 G M_\star\over r_{\rm Ohm}}\right)^{1/2}
={2GM_\star\over \eta},
\ee
which explains why $r_{\rm Ohm}$ is defined as in eq.~(\ref{ohmrad}).}
If we adopt the values
$\eta\approx 2\times 10^{20}$~cm$^2$ s$^{-1}$ and $M_\star \approx
1~M_\odot$ (see Sect.~3), we get $t_{\rm Ohm}\approx 3$~yr, which is
very short compared to the timescale
\be
t_{\rm acc} \equiv  {M_\star \over \dot M} \sim 10^5~{\rm yr},
\label{tacc}
\ee
over which we may expect $M_\star$ or $\Phi_\star$ to have significant
variations. Thus, the quasi-steady approximation is likely to be a good
one.

With these definitions, eq.~(\ref{ss}) becomes
\be
x^2{\partial^2\phi\over\partial x^2}+
x^{3/2}{\partial\phi\over\partial x}=
-(1-\mu^2){\partial^2\phi\over\partial\mu^2}.
\label{ss1}
\ee
Setting $\phi(x,\mu)=F(x)G(\mu)$, we reduce eq.~(\ref{ss1}) by
separation of variables to the couple of ordinary differential equations
\be
(1-\mu^2)G^{\prime\prime}+\Lambda G=0,
\label{ode1}
\ee
{where $\Lambda$ is a separation constant}, and
\be
x^2 F^{\prime\prime}+x^{3/2} F^\prime-\Lambda F=0.
\label{ode2}
\ee

\begin{figure}[t]
\plotone{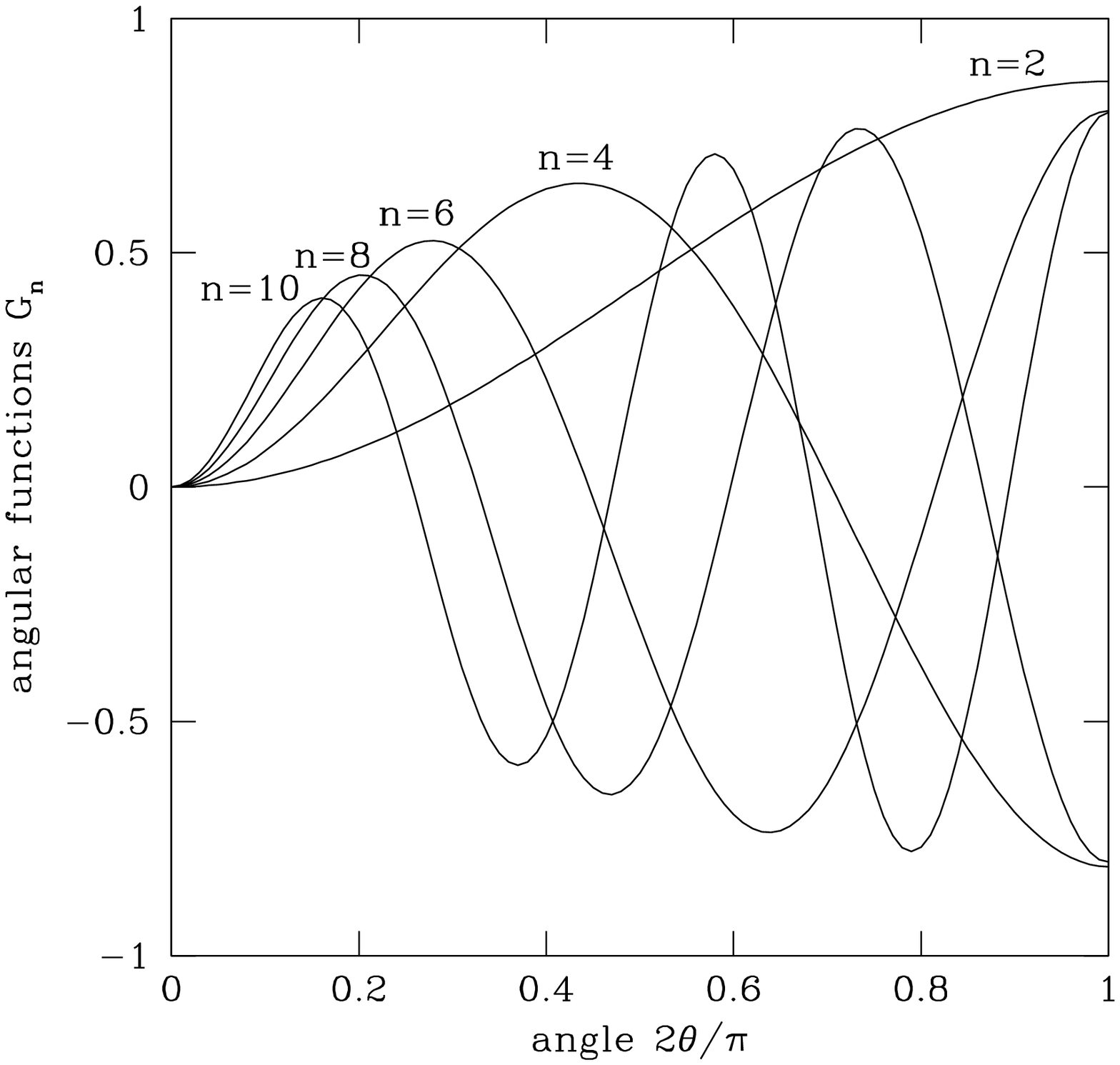}
\caption{The angular functions $G_n(\theta)$ for $n=2$, 4, 6, 8, and 10.}
\label{angular}
\end{figure}

Differentiating eq.~(\ref{ode1}) with respect to $\mu$, and
setting $g(\mu)=G^\prime(\mu)$, we see that $g(\mu)$ satisfies
Legendre's equation,
\be
(1-\mu^2)g^{\prime\prime}-2\mu g^\prime+\Lambda g=0.
\label{leg}
\ee
Solutions of Legendre's equation regular at $\mu=\pm 1$ require
$\Lambda=l(l+1)$, with $l=0,1,2,\ldots$, and are given by Legendre's
polynomials $P_l(\mu)$ of order $l$, i.e. $g(\mu)\propto P_l(\mu)$.
Using the definition of $g$ and eq.~(\ref{ode1}), we then obtain
$G(\mu)\propto (1-\mu^2) P^\prime_l (\mu)$. Therefore, any solution of
eq.~(\ref{ode1}) regular on the polar axis can be written as a linear
combination of polynomials of degree $n=l+1$,
\be
G_n(\mu)=C_n(1-\mu^2) {{\rm d} P_{n-1} \over{\rm d}\mu} =nC_n[\mu
P_{n-1}(\mu)-P_n(\mu)],
\label{defGn}
\ee
where $C_n$ are arbitrary constants. Defining
\be
C_n \equiv \left[{2n-1\over 2n(n-1)}\right]^{1/2},
\label{Cn}
\ee the polynomials $G_n(\mu)$
form an orthonormal set with weight $(1-\mu^2)^{-1}$,
\be
\int_{-1}^{+1}G_n(\mu)G_m(\mu)\;{{\rm d}\mu\over
1-\mu^2}=\delta_{nm}, \label{ortho}
\ee
as can be easily shown by a simple integration by parts and using the
normalization condition of Legendre's polynomials.  Since the flux
function is an even function of $\mu$ over the interval $-1\le\mu\le
1$, the index $n$ takes only even values $n=2,4,\ldots$.
Figure~\ref{angular} shows the functions $G_n(\theta)$ for $n=2$--10.

The boundary condition (\ref{bc}) must be expanded in terms of
$G_n$ polynomials,
\be
\lim_{x\rightarrow\infty}\phi(x,\mu)=
1-|\mu|=\sum_{n=2,4,\ldots}^\infty f_n G_n(\mu),
\ee
{where $f_n$ are the spectral coefficients}. Multiplying both sides
of this equation by $G_m(\mu)/(1-\mu^2)$ and integrating from $\mu=-1$
to $\mu=1$ we easily obtain, using the orthogonality property
(\ref{ortho}), and integrating by parts,
\be
f_n=2C_n\int_0^1 P_{n-1}(\mu)\;{\rm d}\mu=
(-1)^{n/2-1}{2(n-1)!!\over n(n-1)(n-2)!!} C_n,
\label{fn}
\ee
where the latter equality follows from a formula in Byerly (1959).

\begin{figure}[t]
\plotone{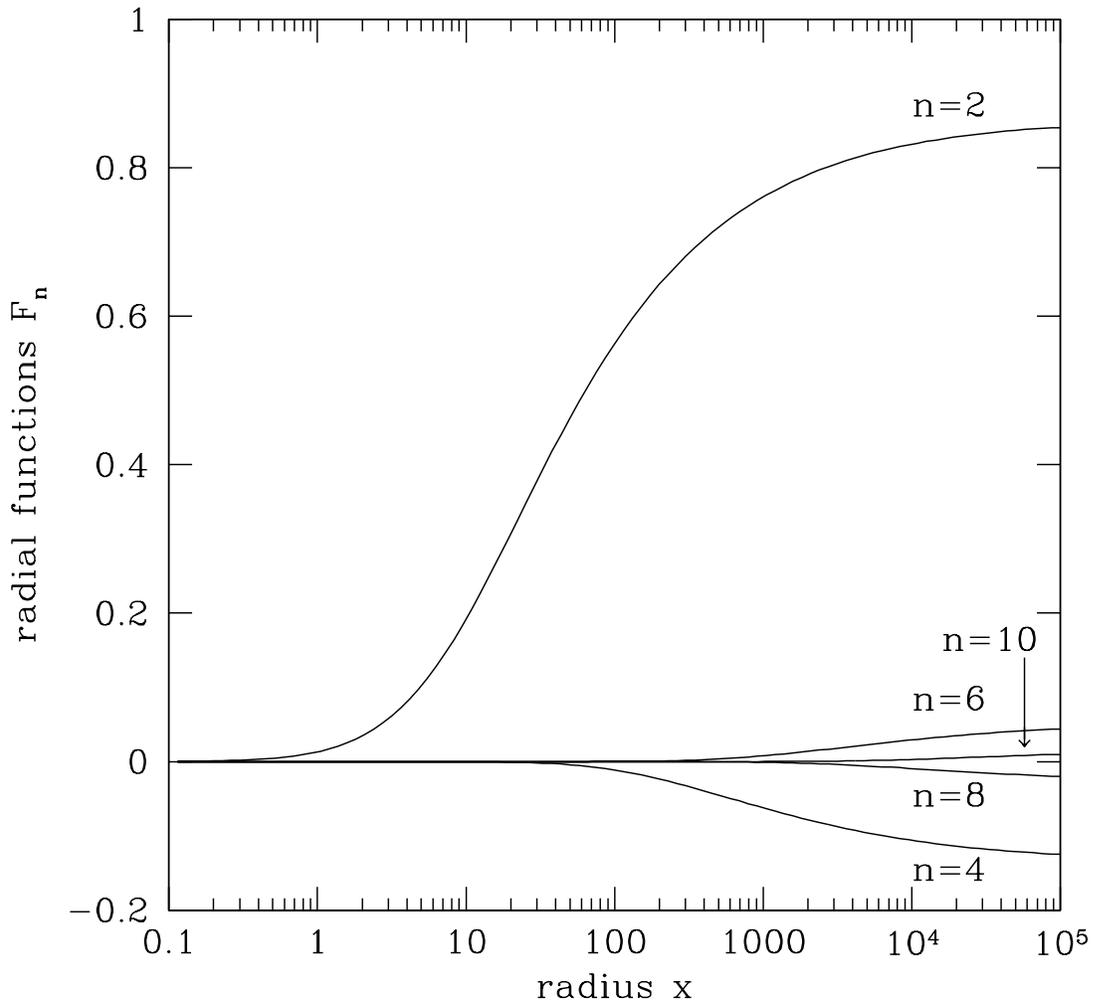}
\caption{The radial functions $F_n(x)$ for $n=2$, 4, 6, 8, and 10.}
\label{radial}
\end{figure}

We now turn to eq.~(\ref{ode2}), with $\Lambda=n(n-1)$.
For each $n$, this equation has to be solved
under the boundary conditions
\be
F_n=0~\mbox{at $x=0$}, \qquad \lim_{x\rightarrow\infty}F_n(x)=f_n,
\label{bc2}
\ee
where the constants $f_n$ are given by eq.~(\ref{fn}).
In terms of the new variables $z$ and $H_n$ defined by
\be
z=2\sqrt{x} \qquad {\rm and} \qquad F_n=z^{2n}e^{-z} H_n,
\ee
eq.~(\ref{ode2}) becomes Kummer's equation (Abramowitz \& Stegun~1965 [13.1.1],
hereafter AS65)
or {\it confluent hypergeometric
equation} (see e.g. Landau \& Lifshitz 1959, Appendix $d$),
\be
zH_n^{\prime\prime}+(b_n-z)H_n^\prime-a_n H_n=0,
\ee
with $a_n=2n-1$ and $b_n=4n-1$.
The general solution of Kummer's equation is (AS65 [13.1.11])
\be
H_n(z)=A_nM(a_n,b_n,z)+B_nU(a_n,b_n,z),
\ee
where $A_n$ and $B_n$ are arbitrary constants, and $M$, $U$ are called
{\it Kummer's functions} or {\it confluent hypergeometric functions} of
the first and second kind, respectively (regular at the origin, and at
infinity, respectively)\footnote{Kummer's equation describes the
radial part of the Coulomb wavefunction.  If $a_n \le 0$, the solutions
of Kummer's equation well-behaved at infinity are associated Laguerre's
polynomials.}. For $z=0$, $M(a_n,b_n,0)=1$ whereas $U(a_n,b_n,z)$
diverges like $z^{1-b_n}=z^{-4n}$.  The latter behavior implies $F_n\sim
x^{-n}$ for small $x$, in contrast with the boundary condition
$F_n(0)=0$. Therefore $B_n=0$ and
\be
F_n(x)=A_n (2\sqrt{x})^{2n}e^{-2\sqrt{x}} M(2n-1,4n-1,2\sqrt{x}).
\ee
The constants $A_n$ are fixed by
imposing the boundary condition at infinity (eq.~\ref{bc2}).
Since $M(a_n,b_n,z)$ has the asymptotic behavior (AS65 [13.1.4])
\be
\lim_{z\rightarrow\infty} M(a_n,b_n,z)=
{(b_n-1)!\over (a_n-1)!}e^z z^{a_n-b_n},
\label{asym}
\ee
we immediately obtain
\be
A_n={(2n-2)!\over (4n-2)!} f_n.
\ee
Figure~\ref{radial} shows the functions $F_n(x)$ for $n=2$--10.

\begin{figure}[t]
\plotone{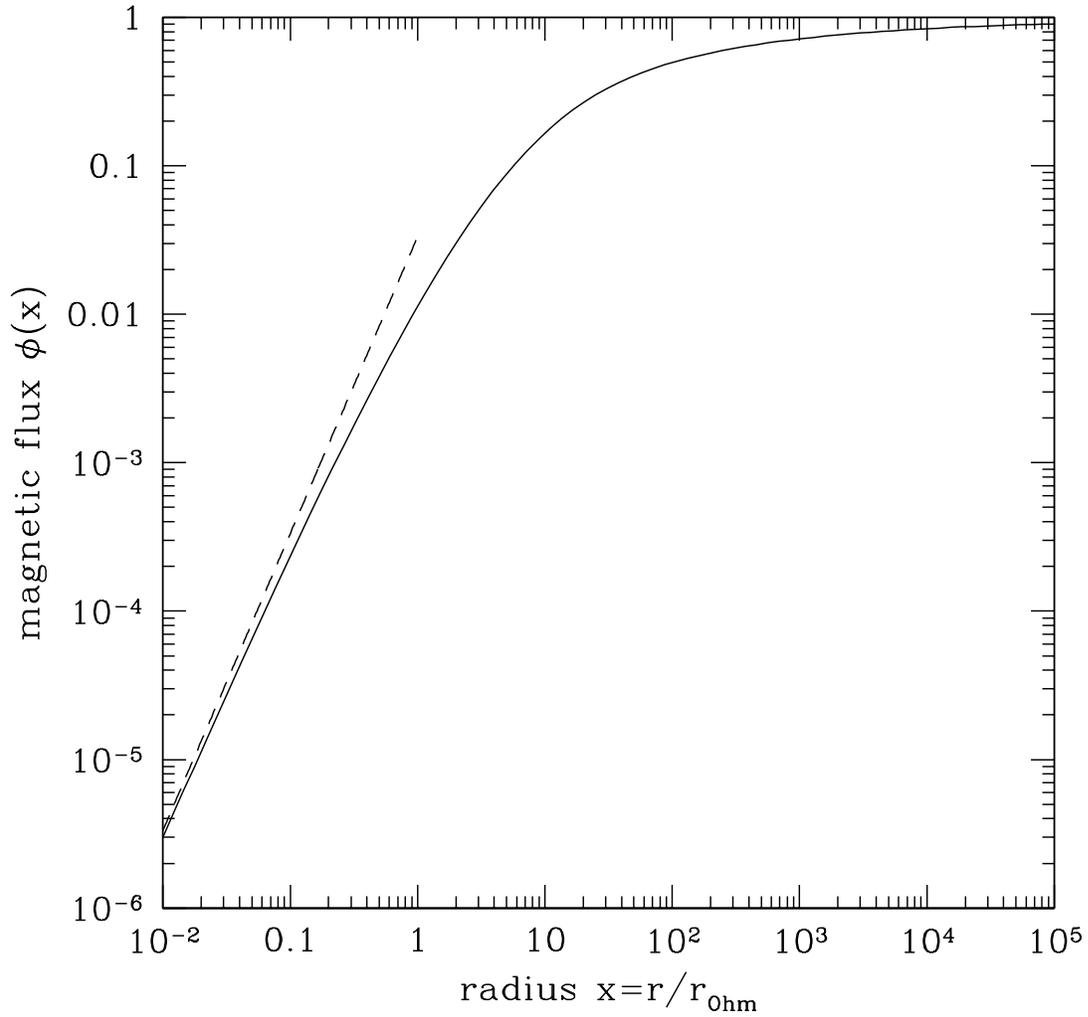}
\caption{The flux function $\phi$ in the midplane ($\theta=\pi/2$) as
function of the distance from the star, in nondimensional units ({\it
solid line}).  The {\it dashed line} shows the uniform field solution
given by eq.~(\ref{unif}) valid for $x\ll1$.}
\label{midplane}
\end{figure}

Combining the angular and radial solutions and summing over 
$n$, we finally obtain
\be
\phi(x,\mu)=e^{-2\sqrt{x}}\sum_{n=2,4,\ldots}^\infty
K_n x^n M(2n-1,4n-1,2\sqrt{x})[\mu P_{n-1}(\mu)-P_n(\mu)]
\label{sol}
\ee
where
\be
K_n=(-1)^{n/2-1}{2^{2n}(n-1)!!(2n-1)!\over (n-2)!!(4n-2)!(n-1)^2n}.
\ee
The function $M(a_n,b_n,z)$ can be evaluated numerically with standard
routines, e.g. the Fortran program {\tt CHGM.FOR} of Zhang \&
Jin~(1996). For $x\gg 1$ is better to use the full asymptotic expansion
eq.~(\ref{asym}), given by AS65 [13.5.1].  For $x \ll 1$, the series in
eq.~(\ref{sol}) is dominated by the $n=2$ term, and we obtain
\be
\lim_{x\rightarrow 0} \phi(x,\theta)={1\over 30} x^2\sin^2\theta,
\label{unif}
\ee
corresponding, as expected, to a uniform magnetic field with vertical
field lines.  Figure~\ref{midplane} shows the flux function $\phi(x,0)$
in the midplane ($\theta=\pi/2$) given by eq.~(\ref{sol}), and its
asymptotic behavior given by eq.~(\ref{unif}). At the Ohm radius the
flux is reduced by a factor $\sim 100$ with respect to the asymptotic
value.

\begin{figure}[t]
\plotone{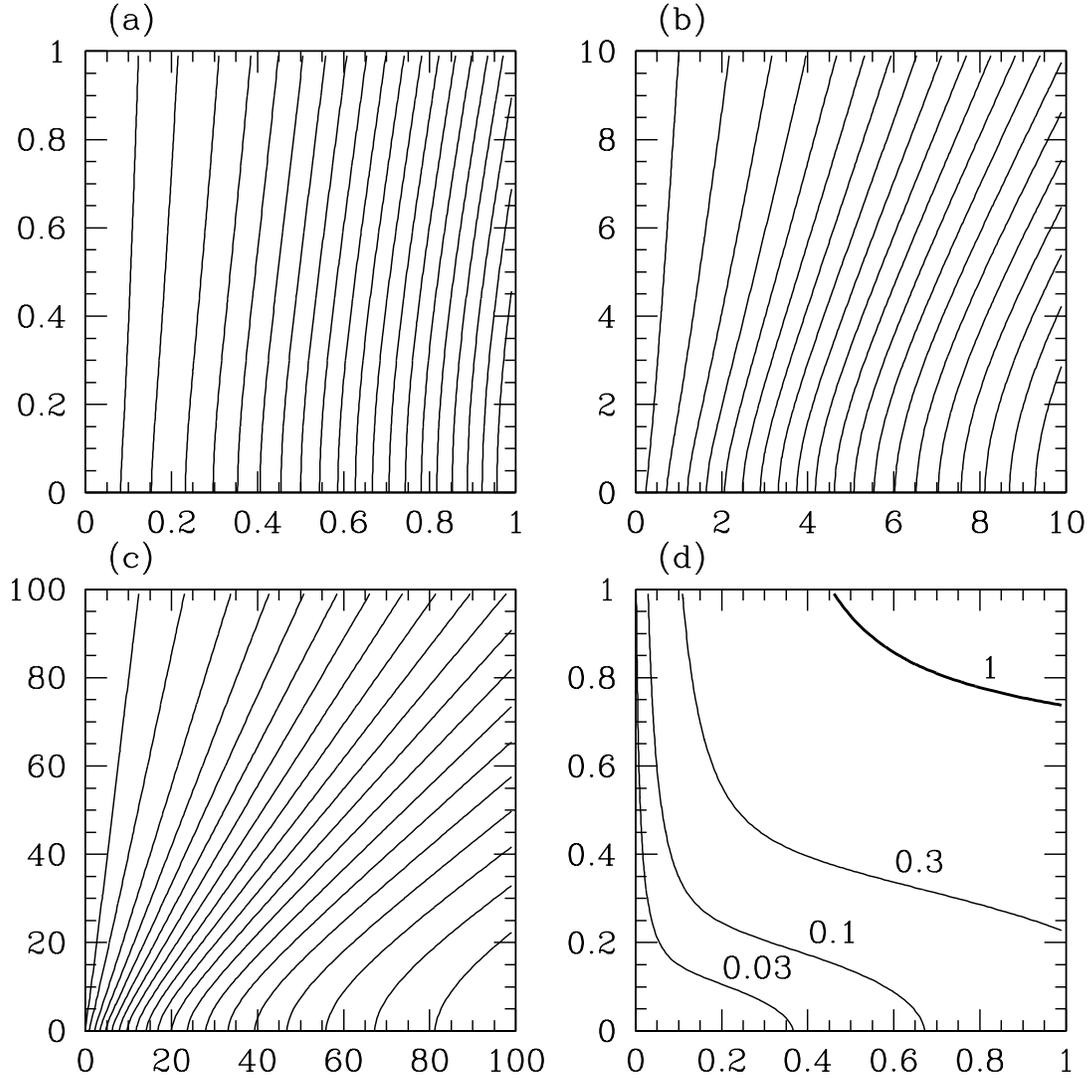}
\caption{({\it a}\/) The nearly uniform magnetic field inside the Ohm
radius, $(\varpi,z)/r_{\rm Ohm} < 1$. The horizontal and vertical axis in
each panel are the cylindrical self-similar coordinates, $\varpi = x
\sin\theta$ and $z=x\cos\theta$.  ({\em b}\/) Same as ({\it a}\/) in
the region $(\varpi,z)/r_{\rm Ohm} < 10$.  ({\it c}\/) Same as ({\em a}\/)
in the region $(\varpi,z)/r_{\rm Ohm} < 100$, showing the asymptotic
convergence to the field of a split monopole at large radii. ({\em d}\/)
Contours of the ratio $|F_L|/|F_g|$ (Lorentz and gravitational forces)
for the density profile corresponding to the collapse of the $H_0=0.5$
toroid (see Paper~I) in the region $(\varpi,z)/r_{\rm Ohm} < 1$. The
kinematic approximation is formally valid in the region below the {\it solid
curve}. The values of the parameters are $M_\star=0.5~M_\odot$, ${\dot
M}=2\times 10^{-5}$~$M_\odot$~yr$^{-1}$, $\lambda_\star=1.7$.}
\label{combo}
\end{figure}

Figure~\ref{combo}$a,b,c$ show the magnetic field lines in the
meridional plane of the collapse region at different scales, computed
according to eq.~(\ref{sol}).  The horizontal and vertical axis in each
panel are the cylindrical self-similar coordinates, $\varpi = x
\sin\theta$ and $z=x\cos\theta$.  Panel ({\em a}\/) shows the nearly
uniform magnetic field inside the Ohm radius, $(R,z)/r_{\rm Ohm} <
1$. In this region the series solution eq.~(\ref{sol}) yields a good
representation with the inclusion of only the first six terms.  Panel
({\em b}\/) shows the magnetic field lines in the region $(R,z)/r_{\rm
Ohm} < 10$.  In this region the series solution eq.~(\ref{sol})
gives a good representation including the first four terms. 
Panel ({\em c}\/) shows
the magnetic field lines in the region $(R,z)/r_{\rm Ohm} < 100$,
showing the asymptotic convergence to the field of a split monopole at
large radii.  In this region the series solution eq.~(\ref{sol})
gives a good representation including the first six terms.

\section{Numerical estimates}

Eq.~(\ref{unif}) implies that the strength of the magnetic field
at the center approaches a constant value given in dimensional
form by
\be
B_c={\Phi_\star\over 30\pi r_{\rm Ohm}^2},
\label{bcenter}
\ee
where $\Phi_\star$ and $r_{\rm Ohm}$ are
given by eq.~(\ref{finf}) and (\ref{ohmrad}), respectively.
Substituting these values, we obtain
\be
B_c={4G^{5/2} M_\star^3\over 15\lambda_\star \eta^4},
\label{bcenter1}
\ee
showing that $B_c$ scales with the electric conductivity and stellar
mass as $\sigma^4 M_\star^3$, a result that may have interesting
consequences for high-mass stars.

To estimate the numerical values of the physical quantities of our
model, we assume a given value for the uniform magnetic field $B_c$
inside the dissipation region, and we derive the remaining quantities in
terms of $B_c$.  An estimate of the intensity of the constant magnetic
field in the central region can be inferred from measurements of
remanent magnetization in meteorites. Observed values range from $\sim
0.1$~G in achondrites to $\sim 1$~G in carbonaceous chondrites,
reaching values up to $\sim 10$~G in chondrules. It is unclear whether
this range of values reflects an actual variation in the strength of
the magnetizing source. Chondrules have randomly oriented
magnetizations which strongly suggest that they record magnetic fields
that predate the accretion of the meteorites, but the measured values
are the most uncertain. Achondrites, on the other hand, have the least
complicated magnetic mineralogies, but local processes such as impacts
may have affected their magnetization history (for reviews, see
Stacey~1976, Levy \& Sonnet~1978, Cisowski \& Hood~1991 and references
therein). Hereafter we assume $B_c\approx 1$~G.

From eq.~(\ref{finf}) and (\ref{bcenter}) we obtain the Ohm radius
\be
r_{\rm Ohm}=\left({G^{1/2}
M_\star\over 15\lambda_\star B_c}\right)^{1/2} \approx
12\;\lambda_\star^{-1/2} \left({M_\star\over M_\odot}\right)^{1/2}
\left({B_c\over 1~\mbox{G}}\right)^{-1/2}~\mbox{AU},
\label{rOhm}
\ee
weakly dependent on the values of the parameters.

We note that magnetic fields of strength $\sim 1$~G, when bent
sufficiently outwards (e.g., by X-winds; Shu et al. 2000), can drive
disk winds (e.g., K\"onigl \& Pudritz 2000).  However, disk winds
driven from footpoints of 2 to 6 AU will have difficulty acquiring the
200 to 300~km~s$^{-1}$ terminal velocities seen in high-speed jet
outflows.  Nevertheless, we cannot rule out, on the basis of these
considerations, the possibility that slow disk winds co-exist with fast
X-winds in YSOs.

The corresponding value for the effective resistivity $\eta$
from the expression for the Ohm radius (eq.~\ref{ohmrad}) is
\be
\eta=\left(4 G^{5/2} M_\star^3
\over 
15\lambda_\star B_c\right)^{1/4}
\approx 2.2\times 10^{20}\;\lambda_\star^{-1/4} 
\left({M_\star\over M_\odot}\right)^{3/4}
\left({B_c\over 1~\mbox{G}}\right)^{-1/4}~\mbox{cm$^2$~s$^{-1}$},
\label{sigma}
\ee
again, weakly dependent on the numerical values of the parameters.
With our fiducial values $\lambda_\star\approx 2$, $M_\star\approx
1$~$M_\odot$, $B_c\approx 1$~G, we obtain $r_{\rm Ohm}\approx 8.5$~AU
and $\eta\approx 2\times 10^{20}$~cm$^2$~s$^{-1}$. {This is larger
by a few orders of magnitude than estimated values from kinetic theory
of the microscopic ohmic resistivity in dense gas and circumstellar
disks (see Sect.~6), suggesting that the dissipation of magnetic flux
probably occurs by some anomalous diffusion process}.

The free-fall velocity at the Ohm radius is $u(r_{\rm Ohm})\approx
14$~km~s$^{-1}$ and the gas density, for an accretion rate $\dot
M\approx 10^{-5}$~$M_\odot$~yr$^{-1}$, is of the order of
$10^9$~cm$^{-3}$. This is lower than the often quoted value of the
density at decoupling of $10^{11}$--$10^{12}$~cm$^{-3}$ (e.g., Nishi,
Nakano, \& Umebayashi 1991), because our adopted effective resistivity is
larger than the conventional electric resistivity.

\section{Joule heating rate}

The magnetic energy annihilated per unit time and unit volume
(Joule heating rate) is
\be
{\eta\over 4\pi} |\nabla \times {\bf
B}|^2= {\eta\over 16\pi^3}\left[{{\cal S}(\Phi)\over
r\sin\theta}\right]^2 ={G M_\ast\over 8\pi^3\eta
r^3\sin^2\theta}\left({\partial\Phi\over\partial r}\right)^2,
\label{dedtdv}
\ee
where we have used eq.~(\ref{ss}) to eliminate ${\cal S}(\Phi)$.
For $x\ll 1$, we can approximate $\phi(x)$ with the asymptotic
expression eq.~(\ref{unif}), obtaining
\be
{1\over x^3\sin^2\theta} \left({\partial \phi\over \partial
x}\right)^2\approx {\sin^2\theta\over 225 x},
\ee
Inserting this
expression in eq.~(\ref{dedtdv}) and integrating over a sphere of
radius $r_{\rm Ohm}$ centered on the origin, we obtain an
approximate estimate of the total energy-dissipation rate
\be
\dot{\cal E}\approx {8G^4 M_\star^5\over 675\lambda_\star^2\eta^5}
\approx 300\lambda_\star^{-2}\left({M_\star\over M_\odot}\right)^5
\left({\eta \over 10^{20}~\mbox{cm$^2$~s$^{-1}$}}\right)^{-5}~L_\odot.
\ee

{The dependence of $\dot{\cal E}$ on the inverse fifth power of
$\eta$ can be easily understood, since} the energy dissipation rate is
proportional to the resistivity times the square of the electric
current density ($\propto \eta^{-12}$) times the volume ($\propto
\eta^6$) in which the current flows (eq.~\ref{dedtdv}). {The
increase of the electric current in the limit of small resistivity (as
$\eta^{-6}$) suggests that the anomalous source of field dissipation
could be associated with current-driven instabilities occurring when the
drift speed of the charged species becomes larger than the ion's
thermal speed, as anticipated by Norman \& Heyvaerts~(1985).} 

With our fiducial values we obtain $\dot {\cal E}\approx 3$~$L_\odot$,
but given the sensitive dependence of $\dot {\cal E}$ on the uncertain
parameter $\eta$ this number may not be very significant.  What is more
interesting is that the adopted resistivity, which is high by
conventional microscopic standards (see Sect.~6), cannot be much lower
without violating observational constraints concerning the total
luminosity from regions of low-mass star formation (see the reviews of
Evans 1999 or Lada \& Lada 2003).  {For example, decreasing the
value of $\eta$ by a factor of 10, increases the central magnetic field
to $B_c\approx 10$~kG, and decreases the Ohm radius to $r_{\rm
Ohm}\approx 0.1 $~AU. Although magnetic fields of kilogauss strength
are measured on the surface of young stars, the energy-dissipation rate
for this reduced value of $\eta$ increases to $\dot{\cal E} \approx 3
\times 10^5$~$L_\odot$, which is unrealistic for solar-mass stars.
Imposing the condition 
that the energy-dissipation rate must be lower than the total
accretion luminosity $GM_\star {\dot M}/R_\star$, we obtain a lower
limit on the resistivity,}
\be
\eta \ga \left(\frac{8G^3M_\star^4R_\star}{675\lambda_\star {\dot M}}\right)^{1/5}
\approx 10^{20} 
\left(\frac{M_\star}{M_\odot}\right)^{4/5}
\left(\frac{R_\star}{R_\odot}\right)^{1/5}
\left(\frac{\dot M}{10^{-5}~\mbox{$M_\odot$~yr$^{-1}$}}
\right)^{-1/5}~\mbox{cm$^2$~s$^{-1}$}.
\ee

Notice also that the energy-dissipation rate $\dot{\cal E}$ scales with
the stellar mass as $M_\star^5$, which is steeper than the
main-sequence luminosity-mass relation, unless the resistivity
increases significantly with increasing $M_\star$. Such a behavior runs
counter to the usual notion that high-mass stars possess more ionizing
potential than low-mass stars. Nevertheless, high mass stars are formed
with mass accretion rates which are more than 100 times larger than
low-mass stars (see, e.g., Osorio, Lizano \& D'Alessio 1999).  In these
conditions of very high circumstellar density the ionization front will
be confined to the stellar surface and the penetrating ionizing agent
for the pseudodisks and disks will not be ultraviolet photons but
cosmic rays and/or X-rays.

\section{Validity of the kinematic approximation}

To check the validity of the kinematic approximation (i.e. the
assumption that the infall velocity is dominated by the gravity of the
central star), we evaluate the ratio of the Lorentz force per unit
volume in the radial direction
\be
F_L={1\over 4\pi}[(\nabla\times {\bf B})\times {\bf B}]_r
={{\cal S}(\Phi)\over 16\pi^3 r^2\sin^2\theta}{\partial\Phi\over\partial r},
\ee
and the gravitational force per unit volume
\be
F_g={GM_\star\rho\over r^2},
\ee
where the density is
\be
\rho={{\dot M}\over 4\pi r^2 |u(r)|} Q(\theta),
\ee
and $Q(\theta)$ yields the flattening of density contours because of
pinching forces associated with the radial magnetic field (see Paper~I).
The ratio of the two forces is
\be
{|F_L|\over |F_g|}=
\lambda_\star^{-2}
\left({t_{\rm acc} \over t_{\rm Ohm} } \right)
\left({\partial\phi\over\partial x}\right)^2
{x\over Q(\theta) \sin^2\theta},
\label{forceratio}
\ee
where we have used eq.~(\ref{ss}) to eliminate ${\cal S}(\Phi)$ and we
used eqs. (\ref{tOhm}) and (\ref{tacc}).  The RHS is proportional to
the ratio of two characteristic times, the accretion time, $t_{\rm acc}
\approx 10^5$ yr, and the crossing time of the ohmic dissipation
region, $t_{\rm Ohm} \approx 3$~yr. The validity of the kinematic
approximation is nevertheless ensured by the fact that the function of
$x$ and $\theta$ on the right of this expression is very small.
Figure~\ref{combo}$d$ shows contours of the ratio $|F_L|/|F_g|$
computed with the function $Q(\theta)$ corresponding to the case
$H_0=1$ of Paper~I. {Notice that the force ratio is low precisely
in the equatorial regions where we might expect the formation of a
centrifugally supported disk to take place if we had included the
effects of angular momentum in the problem.}

We also stress that the ratio of the Lorentz and gravitational forces
given by eq.~(\ref{forceratio}) depends on resistivity as $\eta^{-3}$.
As our solution clearly shows, the non-zero resistivity of the gas
results in a release of the field from the central protostar, as the
gravitational pull of the central star is no longer fully available to
pin the magnetic field of the central regions. However, the released
magnetic field can be too strong for gravity to continue to win over
the Lorentz force.  If $\eta$ is sufficiently large, gravity dominates,
and quasi-steady accretion onto the central source is possible.  If
$\eta$ is too small, magnetic forces overwhelm gravity, and the
accretion region might try to explode outwards. If the coefficient of
resistivity can reach anomalous values, the explosion outwards (when
$\eta$ is small) may be coupled with re-implosion inwards when $\eta$
later becomes (anomalously) large.  It is interesting to speculate that
these alternating reconnection behaviors (``flares'') might correspond
to FU-Orionis outbursts. This mechanism may be an alternate explanation
to the disk thermal instability, possibly aided by protoplanet or
protostellar companions, which has been proposed for the FU-Orionis
phenomenon (e.g., Kawazoe \& Mineshige~1993; Bell et. al~1995; Clarke
\& Syer~1996).

\section{Comparison with other works}

According to calculations by Stepinski~(1992) the electrical
resistivity in the presolar nebula in the range of radii 3--30~AU is in
the range from $\sim 10^{17}$ to $\sim 10^{16}$~cm$^2$~s$^{-1}$ if the
grain size is $\sim 1$~cm, or from $\sim 10^{19}$ to $\sim
10^{16}$~cm$^2$~s$^{-1}$ if the grain size is 0.5~$\mu$m. Clearly, the
physical conditions in a disk are different from those of our infall
model. {Wardle \& Ng~(1999) have evaluated the components of the
conductivity tensor for molecular gas for a variety of grain models, as
function of the gas density. At our fiducial value of $n({\rm H}_2)
\approx 10^9$~cm$^{-3}$, they predict a ohmic resistivity $\eta \approx
10^{16}$~cm$^2$~s$^{-1}$ for a standard grain-size distribution, much
lower than our value $\eta\approx 10^{20}$~cm$^2$~s$^{-1}$.}
Therefore, the work of Stepinski~(1992) {and Wardle \& Ng~(1999)}
suggest that the effective resistivity $\eta$ of the infalling gas had
better be larger than the microscopic electric resistivity, or severe
conflicts with observational data arise.  Given the precedent in solar
physics, this result does not come totally unexpected. Unlike the solar
case, however, the required increase in ``anomalous" resistivity may
not be huge, if the dust grains in the infalling envelopes of
protostars, in contrast with those in protostellar disks, have not
undergone too much growth.

Desch \& Mouschovias~(2001) consider the dissipation of the magnetic
field in a collapsing molecular cloud core, during the pre-pivotal
stage of evolution ($t<0$). Because no central star is present in the
calculations of Desch \& Mouschovias~(2001), the velocity field and the
density profile clearly differ from the free-fall behavior assumed in
this work.  The equation for the evolution of the magnetic field has
the same form as in this work, but Desch \& Mouschovias~(2001) also include
in the resistivity $\eta$ also the contribution of ambipolar diffusion,
\be
\eta=\eta_{\rm Ohm}+\eta_{\rm AD}.
\ee
The latter is given by
\be
\eta_{\rm AD}={|\nabla\Phi|^2\over 16\pi^3\gamma\rho_i\rho_n r^2\sin^2\theta}=
{|{\bf B}|^2\over 4\pi\gamma\rho_i\rho_n},
\ee
where $\rho_i$ and $\rho_n$ are the mass density of ions and neutral,
respectively, and $\gamma$ is the ion-neutral drag coefficient.  The
numerical calculations of Desch \& Mouschovias~(2001) seem to suggest
that the magnetic field approaches a steady-state configuration towards
the end of the run, characterized by a nearly uniform magnetic field of
strength $\approx 0.1$~G over a central region of size $\approx
20$~AU.  Outside this region, the field decreases approximately as
$r^{-1}$, as for field-freezing in a quasi-static isothermal envelope
(Li \& Shu~1996), whereas in our case {of post-pivotal state evolution 
($t>0$)} the field decreases like
$r^{-2}$, as for a split monopole.  In the region of nearly uniform
field (for density larger than $\approx 10^{12}$~cm$^{-3}$), the
ambipolar diffusion resistivity $\eta_{\rm AD}$ reaches $\approx
10^{20}$~cm$^2$~s$^{-1}$, and is larger by one order of magnitude than
the ohmic resistivity (their Fig.~1b). Thus Desch \& Mouschovias~(2001)
conclude that ambipolar diffusion is entirely responsible for the
dissipation (or better redistribution) of the magnetic flux carried by
the infalling gas.  We note, however, that the region of uniform field
in their model contains a negligible mass ($\approx 0.01$~$M_\odot$).
Once the core enters the post-pivotal phase of dynamical collapse,
having dissipated the field in such a tiny fraction of the core's mass
is of little help with respect to solving the magnetic flux problem of
star formation.

Nakano, Nishi, \& Umebayashi~(2002), criticized the results of Desch \&
Mouschovias~(2001) noticing that they neglected the (dominant)
contribution of grains to the ambipolar diffusion resistivity.
Including the grain contribution, Nakano et al.~(2002) found a much
smaller value of the ambipolar diffusion resistivity than Desch \&
Mouschovias~(2001) and concluded that for densities above $\approx
10^{12}$~cm$^{-3}$ the field is dissipated by ohmic resistivity.  To
make progress, future theoretical calculations need to study the
complex relationships among field dissipation, disk formation, and
grain growth, as well as to include self-consistently the back reaction
of the magnetic field and its momentum and energy inputs into the gas
as the field is dissipated.

{An interesting question is the relative
importance of ambipolar diffusion and ohmic dissipation in protostellar
accretion flows. The ratio of the ambipolar diffusion time to the
ohmic dissipation time is given by
\be
{t_{\rm AD}\over t_{\rm Ohm}}= {\eta \over v^2_A \tau_{in}} = 
\eta {4\pi\gamma\rho_i\rho_n \over B^2},
\ee
where $v_A = B/(4 \pi \rho_n)^{1/2}$ is the Alfven speed in the
neutrals, and $\tau_{in} = (\gamma\rho_i)^{-1}$ is the ion-neutral
collision timescale.  Notice that the ratio of timescales depends on
the inverse square of the magnetic field. Thus, as the field is
weakened, electric resistivity will ultimately
dominate the end stages of the process. Indeed, the Ohmic term is the
only one of the three conventionally invoked mechanisms for dissipating
magnetic fields that is linear in ${\bf B}$ in the induction equation
(the so-called Hall term is quadratic and the ambipolar diffusion is
cubic, see e.g. Cowling~1957). This means that Hall and ambipolar
diffusion can never completely annihilate ${\bf B}$; only Ohmic
dissipation can do that.}

Because the induction equation (\ref{ind1}) with only Ohmic
dissipation included is linear in ${\bf B}$, the superposition
principle allows us to modify the approach taken in this paper and
adapt it to other scenarios.  If the process of ambipolar diffusion and
ohmic dissipation occur in series as the gas flows inward,  then, the
compressed core field will be less than the ideal value due to the effect
of ambipolar diffusion in the outer regions.  Adopting the inner limit
of the outer region as the outer limit of our inner calculation,
everything then goes through as before.  One would just have somewhat
different coefficients for the angular functions in eq. (\ref{fn}) and,
more importantly, an overall reduction of the effective flux at
``infinity'', and thus, an increase in $\lambda_\star$. For example, an
increase of the mass-to-flux ratio by ambipolar diffusion by two orders
of magnitude, as found by Tassis \& Mouschovias~(2005a,b), implies
$\lambda_\star \sim  200$.  From equation (\ref{sigma}), the electric
resistivity decreases by a factor of $\sim 4$.  Then, from equation
(\ref{rOhm}), one obtains
$r_{\rm Ohm} \approx 1$~AU. The most important reduction is that the
reconnection luminosity decreases substantially to mean levels
$\dot{\cal E} \approx 0.02 L_\odot$, a small fraction of the bolometric
luminosity, unless the reconnection events occur not steadily but in
infrequent, powerful flares. Finally, since this paper argues that
Ohmic dissipation operates in the innermost region, the annihilation of
the elastic ``core'' that appears as the bottom-most cell in the
calculations of Tassis \& Mouschovias~(2005a,b) may erode the base from
which are launched outwardly propagating shockwaves for their
``spasmodic accretion oscillations''. The point is that the inwardly
advected magnetic flux past the ``last zone'' is not simply
accumulated, as assumed in their calculations, but systematically
destroyed.

\section{Summary and conclusions}

Assuming quasi-steady state and a spatially uniform resisivity
coefficient $\eta$, we have solved the problem of magnetic field
dissipation during the accretion phase of star formation. We have
adopted the velocity field determined in a previous study of the
gravitational collapse of a magnetized cloud (Galli et al.~2006), and
we have ignored the back reaction of the changed magnetic topology on
the flow (using the field-freezing calculations to provide what we have
called a ``kinematic approximation''). With these assumptions, we have
solved the problem analytically, and checked a posteriori the validity
of our approximations.  According to our solution, the magnetic field
morphology changes from radial at large distances to asymptotically
uniform approaching the origin, so that the magnetic flux accreted by
the central star is zero at any time. 

To determine the value of the resistivity coefficient $\eta$ we have
considered the restrictions imposed by measurements of magnetic fields
in meteorites. These constraints require an effective resistivity $\eta
\approx 10^{20}$~cm$^2$~s$^{-1}$, probably several orders of magnitude
larger than the microscopic electric resistivity of the infalling gas.
Having shown that ohmic resistivity can dissipate enough magnetic field
to solve the magnetic flux problem satisfying the available
observational constraints, one now needs to solve the full dynamic
problem of magnetic field dissipation and formation of a centrifugally
supported protoplanetary disk in a self-consistent way.

\acknowledgements

DG and SL acknowledge financial support from the Theoretical Institute
for Advance Research in Astrophysics (TIARA), CRyA/UNAM, DGAPA/UNAM and
CONACyT (Mexico), and INAF-Osservatorio Astrofisico di Arcetri (Italy),
where parts of the research presented in this paper were done. These
authors are also grateful for the warm hospitality of members and staff
of these institutions.  The research work of FS and MC in Taiwan is
supported by the grant NSC92-2112-M-001-062.

\clearpage

\clearpage

\clearpage

\clearpage

\end{document}